\documentclass[english]{article}
\usepackage[T1]{fontenc}
\usepackage[latin9]{inputenc}
\usepackage{amstext}

\makeatletter

\DeclareRobustCommand{\greektext}{%
  \fontencoding{LGR}\selectfont\def\encodingdefault{LGR}}
\DeclareRobustCommand{\textgreek}[1]{\leavevmode{\greektext #1}}
\DeclareFontEncoding{LGR}{}{}
\DeclareTextSymbol{\~}{LGR}{126}

\newcommand{\lyxaddress}[1]{
\par {\raggedright #1
\vspace{1.4em}
\noindent\par}
}

\usepackage[samesize]{cancel}

\makeatother

\usepackage{babel}
\begin{document}

\title{Light Cone Reflection and the Spectrum of Neutrinos}

\author{Alan Chodos%
\thanks{chodos@aps.org%
}}

\maketitle

\lyxaddress{
\[
American\: Physical\: Society,\: College\: Park,\: MD\:20740
\]
}
\begin{abstract}
We extend the treatment of neutrinos within the context of SIM(2)
Very Special Relativity (VSR) by adding a new discrete symmetry that
we call Light Cone Reflection (LCR). We construct a Lagrangian that
exhibits both VSR and LCR symmetry, and find that the spectrum involves
two neutrinos, one tachyonic and the other not, with the same absolute
value of the mass parameter. We argue that LCR symmetry offers a rationale
for introducing tachyonic neutrinos.
\end{abstract}

\section*{Introduction}

During the recent flurry of interest in tachyonic neutrinos occasioned
by the rise\cite{OPERA} and fall\cite{ICARUS} of the result from
the OPERA experiment, it became clear that although the idea of their
existence had been proposed long ago\cite{Cawley,CHK}, there was
no compelling theoretical reason behind it. Contrariwise, there were
powerful theoretical arguments, most notably those in reference \cite{Cohen/Glashow},
that seriously questioned whether OPERA had even seen what it claimed
to see. 

Notwithstanding the climate of doubt surrounding the OPERA result,
it is still possible that at least one of the neutrino mass eigenstates
is tachyonic, although the effect in a time-of-flight measurement
will necessarily be less than the parts per 10$^{5}$ that OPERA reported.
In anticipation of future experiments, possibly with renewed claims
of a positive signal, one of the motivations of this paper is to offer
a theoretical rationale for the existence of tachyonic neutrinos.

In their 1985 paper, the authors of ref. \cite{CHK} proposed the
equation

\[
\left(i\cancel{\nabla}\gamma^{5}-m\right)\psi=0
\]
to describe tachyonic neutrinos, in the hope that a fully Lorentz-invariant
theory could be constructed. However, a quantum theory based on this
equation led to difficulties that have not been fully overcome. Nor
did the theory gain any new symmetry or other desirable features as
a result of the properties of this equation.

Meanwhile, the idea that Lorentz symmetry might be broken began to
gain some traction\cite{Coleman/Glashow}, and, in particular, Colladay
and Kostelecký introduced\cite{SME1} the Standard Model Extension,
which systematized the search for Lorentz-violating effects. Recently,
Kostelecký and Mewes\cite{SME2} used the SME to provide a detailed
discussion of Lorentz violating effects in the neutrino sector.

In 2006, Cohen and Glashow\cite{CG1} introduced a form of Lorentz
violation, called Very Special Relativity (VSR), which proposes that
the relativistic symmetry is one of four particular subgroups of the
Lorentz Group. CG argue that almost all the well-known consequences
of Lorentz invariance follow from VSR as well. In addition, VSR requires
that the P, T and CP symmetries all be broken. Imposing any one of
them elevates the symmetry to the full Lorentz Group.

In a second paper\cite{CG2}, CG concentrate on the largest of the
four subgroups, the maximal subgroup SIM(2), which is a four parameter
group. They study neutrinos in this context, pointing out that it
is possible to have neutrino masses that neither violate lepton number
nor require the existence of sterile neutrinos, and positing a non-local
wave equation for the neutrino.

To implement the SIM(2) version of VSR, one introduces a null four-vector,
$n^{\mu}$, such that the physics can depend on the spatial direction
of $n$ but not its magnitude. If we label this direction to be the
z axis, then $n^{\mu}$ has the form $n^{\mu}=a(1,0,0,1)$, where
$a$ is an arbitrary non-zero real number on which the physics cannot
depend. We shall therefore set $a=1$. 

It is the purpose of this paper to point out that, although VSR reduces
the symmetry compared to normal Special Relativity, the existence
of $n^{\mu}$ allows us to enlarge the symmetry in a different way,
by introducing a new, discrete, spacetime symmetry that lies outside
the Lorentz Group. We call this symmetry Light Cone Reflection, or
LCR.

In the following sections, we define LCR, derive some of its properties,
and relate it to a new one-parameter family of spacetime transformations.
Then, building on a Lagrangian introduced by Álvarez and Vidal\cite{Alvarez},
we construct a theory of neutrinos that is both VSR and LCR invariant,
and discuss some of its features.

\section*{LCR and its Continuous Generalization}

Consider the transformation

\[
x^{\mu}\rightarrow y^{\text{\textgreek{m} }}=x^{\text{\textgreek{m} }}-n^{\text{\textgreek{m}}}\frac{x^{2}}{n\text{·\ensuremath{x}}}
\]
where $n^{\text{\textgreek{m}}}$ is the null vector associated with
the SIM(2) symmetry. Note that the transformation is independent of
the overall scale of $n^{\text{\textgreek{m}}}$, as required. We
call this \textquotedblleft{}Light Cone Reflection\textquotedblright{},
principally because 

\[
y^{2}=-x^{2}
\]
i.e. LCR maps timelike vectors into spacelike ones, and vice-versa.
Also, we see that

\[
n\text{·}y=n\text{·}x
\]
and

\[
x^{\text{\textgreek{m}}}=y^{\text{\textgreek{m}}}-n^{\text{\textgreek{m}}}\frac{y^{2}}{n\text{·}y}
\]
i.e if we repeat the transformation we get back where we started.
In addition, 

\[
x\text{·}y=0
\]
This transformation is defined for all $x$ such that $n\text{·}x\neq0$
. Since $n$ is null, $n\text{·}x$ cannot vanish for any timelike
$x$, and for lightlike $x$, it vanishes only for $x$ proportional
to $n$. There is, however, a subset of spacelike $x$ for which $n\text{·}x$
vanishes. In some sense, there are \textquotedblleft{}more\textquotedblright{}
spacelike than timelike vectors, so an invertible mapping between
the two kinds of vectors cannot include all the spacelike vectors. 

A simple calculation yields 

\[
\frac{\text{\ensuremath{\partial}}}{\text{\ensuremath{\partial}\ensuremath{\ensuremath{y^{\text{\textgreek{n}}}}}}}=\frac{\text{\ensuremath{\partial}}}{\text{\ensuremath{\partial}\ensuremath{\ensuremath{x^{\text{\textgreek{n}}}}}}}-\frac{\text{\ensuremath{\partial}}}{\text{\ensuremath{\partial}\ensuremath{\ensuremath{x^{\text{\textgreek{n}}}}}}}\left[\frac{x^{2}}{n\text{\textgreek{;}}x}\right]n^{\text{\textgreek{m}}}\frac{\text{\ensuremath{\partial}}}{\text{\ensuremath{\partial}\ensuremath{\ensuremath{x^{\text{\textgreek{m}}}}}}}.
\]
 Observing that

\[
n\text{·}\frac{\text{\ensuremath{\partial}}}{\text{\ensuremath{\partial}x}}\left[\frac{x^{2}}{n\text{\textgreek{;}}x}\right]=2,
\]
we see that 

\[
n\text{·}\frac{\text{\ensuremath{\partial}}}{\text{\ensuremath{\partial}y}}=-n\text{·}\frac{\text{\ensuremath{\partial}}}{\text{\ensuremath{\partial}x}}.
\]
This property will play a key role in constructing a VSR- and LCR-invariant
theory in the next section. 

We close this section with a brief discussion of a one-parameter group
of transformations that includes LCR. It is convenient to write the
transformation in the form 

\[
y^{\mu}\left(\alpha\right)=x^{\mu}-\frac{1}{2}\left(1-\alpha\right)n^{\mu}\frac{x^{2}}{n\text{·}x}
\]
Written this way, the product of a transformation parameterized by
$\alpha$ and one parameterized by $\beta$ is simply given by a transformation
parameterized by the product $\alpha\beta$. The identity is $\alpha=1$,
LCR is $\alpha=-1$, and the value $\alpha=0$ must be excluded because
it doesn\textquoteright{}t have an inverse. 

The earlier formulas generalize easily to accommodate $\alpha$. We
have: 

\[
y^{2}=\alpha x^{2};
\]

\[
n\text{·}y=n\text{·}x;
\]

\[
x^{\mu}=y^{\mu}-\frac{1}{2}\left(1-\frac{1}{\alpha}\right)n^{\mu}\frac{y^{2}}{n\text{·}y};
\]

\[
x\text{·}y=\frac{1}{2}\left(1+\alpha\right)x^{2};
\]

\[
\frac{\partial}{\partial y^{\nu}}=\frac{\partial}{\partial x^{\nu}}-\frac{1}{2}\left(1-\frac{1}{\alpha}\right)\frac{\partial}{\partial x^{\nu}}\left[\frac{x^{2}}{n\cdot x}\right]n^{\mu}\frac{\partial}{\partial x^{\mu}};
\]
and

\[
n\cdot\frac{\partial}{\partial y}=\frac{1}{\alpha}n\cdot\frac{\partial}{\partial x}.
\]

Finally, we observe that

\[
\frac{\partial y^{\mu}}{\partial x^{\nu}}=\delta_{\nu}^{\mu}-\frac{1}{2}\left(1-\alpha\right)n^{\mu}\frac{\partial}{\partial x^{\nu}}\left[\frac{x^{2}}{n\cdot x}\right],
\]
and, with $n^{\mu}=\left(1,0,0,1\right),$we can explicitly evaluate
the Jacobean of the transformation, 

\[
\left|\frac{\partial y^{\mu}}{\partial x^{\nu}}\right|=|\alpha|.
\]

\section*{VSR- and LCR-invariant Lagrangian}

Cohen and Glashow postulate the following VSR-invariant non-local
wave equation for the neutrino:

\[
\left(\cancel{p}-\frac{1}{2}\frac{m^{2}}{n\text{·}p}\cancel{n}\right)\nu=0
\]
 which implies the dispersion formula $p^{2}=m^{2}.$

Álvarez and Vidal \cite{Alvarez} exhibited a Lagrangian density that
yields the CG equation:

\[
L=i\bar{\nu}\cancel{\partial}\nu+i\bar{\chi}n\text{·}\partial\psi+i\bar{\psi}n\text{·}\partial\chi+\frac{i}{2}m[\bar{\chi}\cancel{n}\nu+\bar{\psi}\nu-\bar{\nu}\cancel{n}\chi-\bar{\nu}\psi]
\]
When $n^{\mu}$ is scaled by a VSR transformation, $\chi$ must be
scaled oppositely to compensate.

Let us generalize the Lagrangian slightly:

$ $
\[
L=i\bar{\nu}\cancel{\partial}\nu+i\bar{\chi}n\text{·}\partial\psi+i\bar{\psi}n\text{·}\partial\chi+\frac{i}{2}[m_{1}\bar{\chi}\cancel{n}\nu+m_{2}\bar{\psi}\nu-m_{1}^{*}\bar{\nu}\cancel{n}\chi-m_{2}^{*}\bar{\nu}\psi]
\]
This still yields the CG equation, but now with $m^{2}$ replaced
by Re$(m_{2}^{*}m_{1})$.

When we make an LCR transformation, we have two things to worry about:
(i) the $\bar{\nu}\cancel{\partial}\nu$ term is not invariant, because 

\[
\partial_{\mu}\rightarrow\partial_{\mu}-\partial_{\mu}\left[\frac{x^{2}}{n\text{·}x}\right]n\text{·}\partial
\]
and (ii) the terms $i\bar{\chi}n\text{·}\partial\psi+i\bar{\psi}n\text{·}\partial\chi$
change sign because $n\text{·}\partial\rightarrow-n\text{·}\partial$
.

We consider the second problem first. We can compensate for the sign
change by changing the sign of either $\chi$ or $\psi$ (but not
both). Let us choose $\psi$. Then the mass terms proportional to
m$_{2}$ and m$_{2}{}^{*}$ change sign. To fix this, we relabel $\nu$
as $\nu_{1}$ and introduce a second neutrino field $\nu_{2}$, with
opposite sign of the $m_{2}$ mass terms compared to $\nu_{1}$. The
Lagrangian then becomes 

\begin{eqnarray*}
L & = & i\bar{\nu_{1}}\cancel{\partial}\nu_{1}+i\bar{\nu_{2}}\cancel{\partial}\nu_{2}+i\bar{\chi}n\text{·}\partial\psi+i\bar{\psi}n\text{·}\partial\chi+\frac{i}{2}[m_{1}\bar{\chi}\cancel{n}\nu_{1}+m_{2}\bar{\psi}\nu_{1}-m_{1}^{*}\bar{\nu_{1}}\cancel{n}\chi-m_{2}^{*}\bar{\nu_{1}}\psi\\
 &  & +m_{1}\bar{\chi}\cancel{n}\nu_{2}-m_{2}\bar{\psi}\nu_{2}-m_{1}^{*}\bar{\nu_{2}}\cancel{n}\chi+m_{2}^{*}\bar{\nu_{2}}\psi]\\
\end{eqnarray*}
 Under LCR, $n\text{·}\partial\rightarrow-n\text{·}\partial,$ $\chi\rightarrow\chi,$
$\psi\rightarrow-\psi,$ $and$ $\nu_{1}\longleftrightarrow\nu_{2}$
, leaving $L$ invariant (except for the $\cancel{\partial}$ terms,
which we will treat below).

Variation of this Lagrangian yields coupled equations for $\nu_{1}$
and $\nu_{2}$ which, when diagonalized, lead to a pair of CG equations,
one with $m^{2}$ replaced by $\left|m_{1}m_{2}\right|$ and the other
with $m^{2}$ replaced by $-\left|m_{1}m_{2}\right|$ ; i.e. the spectrum
contains a pair of neutrinos, one a normal particle, and the other
a tachyon.

To deal with the non-invariance of $\bar{\nu}\cancel{\partial}\nu$,
we introduce a vector field $A_{\mu}$ by replacing the derivative
$\partial_{\mu}$ in $\cancel{\partial}$ with a covariant derivative
$D_{\mu}$:

\[
\partial_{\mu}\rightarrow\partial_{\mu}-\frac{1}{M}A_{\mu}n\text{·}\partial\equiv D_{\mu}.
\]
(Although we call the new field $A_{\mu}$, we do not mean to suggest
that it is related to the photon.) We need to choose the transformation
of $A_{\mu}$ so that $D_{\mu}$ is invariant under LCR. Let $V(x)\equiv\frac{x^{2}}{n\text{·}x}$
. Then $n\text{·}\partial V=2.$ $ $We have

\[
\partial_{\mu}^{'}=\partial_{\mu}-\partial_{\mu}Vn\text{·}\partial
\]
 and $ $
\[
n\text{·}\partial^{'}=-n\text{·}\partial
\]
 so $ $
\[
D_{\mu}^{'}=\partial_{\mu}-\partial_{\mu}Vn\text{·}\partial+\frac{1}{M}A_{\mu}^{'}n\text{·}\partial.
\]
Therefore, with 
\[
A_{\mu}^{'}=-A_{\mu}+M\partial_{\mu}V
\]
 we have $D_{\mu}^{'}=D_{\mu},$ as required. The appearance of the
dimensionful parameter M in the coupling of $A_{\mu}$ means that
the interaction is non-renormalizable, and suggests that this Lagrangian
will give way to new physics at energy scales $\sim M.$ 

In addition to replacing $\partial_{\mu}$ with $D_{\mu},$ for each
$\nu_{i}$ we must also add the term 

\[
\frac{-i}{2M}\bar{\nu}\gamma^{\mu}\nu\, n\text{·}\partial A_{\mu}
\]
to L, in order to preserve its Hermiticity.

Having introduced $A_{\mu},$ we need to find a suitable kinetic term
that is both VSR and LCR invariant. This turns out to be a little
messy. Standard operating procedure dictates that we evaluate

\[
\left[D_{\mu},\: D_{\nu}\right]=-\frac{1}{M}F_{\mu\nu}n\text{·}\partial,
\]
 which yields

\[
F=\partial_{\mu}A_{\nu}-\partial_{\nu}A_{\mu}-\frac{1}{M}\left(A_{\mu}n\text{·}\partial A_{\nu}-A_{\nu}n\text{·}\partial A_{\mu}\right).
\]
 It is easily verified that, under LCR, $F_{\mu\nu}\rightarrow-F_{\mu\nu}.$
One cannot, however, simply use $-\frac{1}{4}F_{\mu\nu}F^{\mu\nu}$
as the kinetic term for $A_{\mu}$. We see from the definition of
D$_{\mu}$ that scaling $n^{\mu}$ under VSR must be accompanied by
an inverse scaling of $A_{\mu}.$ Therefore $ $$-\frac{1}{4}F_{\mu\nu}F^{\mu\nu}$
is not VSR invariant. Each factor of F must be accompanied by an additional
factor of n. We define 

\[
J_{\nu}=n^{\mu}F_{\mu\nu}.
\]
 Then $J^{\mu}J_{\mu}$ has the necessary symmetry. However, let us
define the additional quantities 

\[
Q_{\nu}=n\text{·}\partial A_{\nu},
\]
 and 
\[
K_{\nu}=\partial_{\nu}\left(n\text{·}A\right)-\frac{1}{M}A_{\nu}n\text{·}\partial\left(n\text{·}A\right).
\]
 We observe that, under LCR, $Q_{\nu}\rightarrow Q_{\nu},$ and $K_{\nu}\rightarrow-K_{\nu},$
and

\[
J_{\nu}=Q_{\nu}-K_{\nu}-\frac{1}{M}\left(n\text{·}A\right)Q_{\nu}.
\]
 Thus from invariance requirements alone, a suitable kinetic term
would be $a_{1}J_{\nu}J^{\nu}+a_{2}K_{\nu}K^{\nu}+a_{3}Q_{\nu}Q^{\nu}$,
where the $a_{i}$ are arbitrary coefficients. It might be possible
to impose the additional LCR-invariant constraint $Q_{\nu}=0$, in
which case, we would have $J_{\nu}=-K$$_{\nu},$ and the remaining
kinetic term would be essentially unique.

\section*{Conclusions}

This paper is being written at a moment when the idea of tachyonic
neutrinos, never one of great popularity, is at a particularly low
ebb. Despite interesting theoretical and phenomenological contributions
over the years, for example references \cite{Radzikowski,Rembielinski,Nuclear Null Tests,Recami,Schwartz,Ehrlich,Jentschura},
this has always been the pursuit of a small minority, and now the
majority has been convinced, by the retraction of the OPERA result
and the report of some new experiments\cite{ICARUS}, that neutrinos
have been shown not to travel faster than light.

Of course this is not what has been shown. OPERA claimed, in a time
of flight measurement of neutrinos whose energies averaged 17 GeV,
that the speed exceeded that of light by some parts in 10$^{5}.$
This is an extremely awkward result, because, assuming that the superluminal
neutrinos obey a Lorentz-invariant dispersion formula, the mass parameter
has to be in the range of hundreds of MeV. The demise of this result
is not cause for concern among aficionados of the tachyonic neutrino
hypothesis. Assuming that the neutrinos have mass parameters in the
meV range, the deviation from light speed in the OPERA experiment,
or in any of the other time of flight measurements so far reported,
would be unmeasurably small.

Theoretical papers on tachyonic neutrinos, and on tachyons more generally,
have mainly been concerned with the question of whether a sensible
theory involving superluminal particles can be formulated. The difficulties
in this regard are quite formidable, and despite a number of careful
and ingenious approaches, there is so far no generally accepted solution.
The purpose of this paper is somewhat different. We want to make the
argument not that tachyons are possible, but that, if introduced appropriately,
they can be viewed as the necessary concomitant of a new symmetry
of nature. We have exhibited such a symmetry, constructed an invariant
Lagrangian, and shown that tachyons and non-tachyons arise symmetrically
in its spectrum.

At best our Lagrangian is only an effective one; the derivative couplings
that we have introduced probably mean that it is non-renormalizable.
Many open questions remain, among them whether an LCR-invariant theory
can be successfully quantized, whether it can be incorporated into
an extension of the Standard Model, and whether it can be extended
to the continuous symmetry that we discussed briefly above. An obstacle
to the latter is the factor of $\left|\alpha\right|$ contributed
by the Jacobean, compensating for which probably requires the addition
of at least one more field.

Finally we note that, in VSR alone, one could contemplate restricting
explicit deviations from full Lorentz symmetry to the neutrino sector,
with leakage to the rest of the Standard Model being quite small.
In our extension to LCR, we have introduced the gauge-like field $A_{\mu},$
which must accompany all derivatives (except those in the form $n\text{·}\partial$)
$ $ in the full Lagrangian in order to maintain the symmetry. The
effects of $A_{\mu}$ are suppressed by the factor $\frac{1}{M}$,
but the consequences of imposing LCR will inevitably be more widespread
than those of VSR alone.

\section*{Acknowledgments}

I have benefited from conversations with Alan Kostelecký and Gregg
Gallatin. I am grateful to Fred Cooper for collaboration on earlier,
unpublished work on tachyonic neutrinos. I have enjoyed stimulating
correspondence with Marek Radzikowski. Thanks to the Indiana University
Center for Spacetime Symmetries for hospitality at a workshop.


\begin{thebibliography}{10}
\bibitem{OPERA}T. Adam et al. {[}OPERA Collaboration{]}, arXiv:1109.4897v2

\bibitem{ICARUS}M. Antonello et al. {[}ICARUS Collaboration{]}, arXiv:1203.3433;
N. Yu. Agafonova et al. {[}LVD and OPERA Collaborations{]}, arXiv:1206.2488.

\bibitem{Cawley}R. G. Cawley, Bull. Am. Phys. Soc. \textbf{15}, 25
(1970), and Lettere al Nuovo Cimento \textbf{3}, 523 (1972).

\bibitem{CHK}A. Chodos, A. Hauser and V. A. Kostelecký, Phys. Lett.
\textbf{150B}, 431 (1985).

\bibitem{Cohen/Glashow}A.G. Cohen and S.L. Glashow, Phys. Rev. Lett.
\textbf{107}, 181803 (2011).

\bibitem{Coleman/Glashow} A. Kostelecký and R. Potting, Phys. Rev.
D \textbf{51}, 3923 (1995); S. Coleman and S. L. Glashow, Phys. Lett.
\textbf{B405}, 249 (1997) and Phys. Rev. D\textbf{ 59}, 116008 (1999).

\bibitem{SME1} D. Colladay and A. Kostelecký, Phys. Rev. D \textbf{55},
6760 (1997) and\textbf{ 58}, 116002 (1998).

\bibitem{SME2} A. Kostelecký and M. Mewes, Phys. Rev. D\textbf{ 85},
096005 (2012).

\bibitem{CG1}A. G. Cohen and S. L. Glashow, Phys. Rev. Lett. \textbf{97},
021601 (2006).

\bibitem{CG2} A. G. Cohen and S. L. Glashow, arXiv:hep-ph/0605036v1.

\bibitem{Alvarez} Enrique Álvarez and Roberto Vidal, Phys. Rev. D\textbf{
77}, 127702 (2008).

\bibitem{Radzikowski}Marek J. Radzikowski, arXiv:0801.1957, 0804.4534,
and 1007.5418.

\bibitem{Rembielinski}Jacek Ciborowski and Jakub Rembielinski, Eur.
Phys. J. C\textbf{ 8}, 157 (1999).

\bibitem{Nuclear Null Tests}A. Chodos et al., Mod. Phys. Lett. A
\textbf{7}, 467 (1992); A. Chodos and A. Kostelecký, Phys. Lett. \textbf{B336},
295 (1994).

\bibitem{Recami}E. Giannetto, G.D. Maccarrone, R. Mignani and E.
Recami, Phys. Lett. \textbf{B178}, 115 (1986).

\bibitem{Schwartz}C. Schwartz, J. Math. Phys. \textbf{52}, 052501
(2011), especially Appendix B.

\bibitem{Ehrlich} R. Ehrlich, arXiv:1110.0736, 1111.0502, and 1204.0484.

\bibitem{Jentschura}U. D. Jentschura and B. J. Wundt, arXiv:1110.4171,
1205.0521, and 1206.6342.\end{thebibliography}
\end{document}